\documentclass[journal]{IEEEtran}

\makeatletter
\let\NAT@parse\undefined
\makeatother

\usepackage[numbers,sort&compress]{natbib}
   
\usepackage{amssymb}
\usepackage[cmex10]{amsmath}
\usepackage{mathtools}
\usepackage{multirow}
\usepackage{flushend}
\usepackage[final]{graphicx}

\newtheorem{lemma}{Lemma}
\newtheorem{definition}{Definition}
\usepackage{color}

\def \E {\mathbb E}

\makeatletter
\def\blfootnote{\xdef\@thefnmark{}\@footnotetext}
\makeatother

\title{The $\alpha$-$\kappa$-$\mu$ Shadowed Fading Distribution: Statistical Characterization and Applications}

\author{\IEEEauthorblockN{Pablo Ramirez-Espinosa\IEEEauthorrefmark{1}, Jules M. Moualeu\IEEEauthorrefmark{2}, Daniel Benevides da Costa\IEEEauthorrefmark{3} and F. Javier Lopez-Martinez\IEEEauthorrefmark{1}}
\IEEEauthorblockA{\IEEEauthorrefmark{1}Dpto. de Ingenier\'\i a de Comunicaciones, Universidad de Malaga, 29071, Malaga, Spain\\\IEEEauthorrefmark{2}School of Electrical and Information Engineering, University of the Witwatersrand, Johannesburg, South Africa\\ \IEEEauthorrefmark{3} Department of Computer Engineering, Federal University of Ceara, Sobral, CE, Brazil}\\ Email: pre@ic.uma.es, jules.moualeu@wits.ac.za, danielbcosta@ieee.org, fjlopezm@ic.uma.es.
}

\begin{document}
\maketitle

\begin{abstract}
\blfootnote{This work has been submitted to the IEEE for possible publication. Copyright may be transferred without notice, after which this version may no longer be accessible.}We introduce the $\alpha$-$\kappa$-$\mu$ shadowed ($\alpha$-KMS) fading distribution as a natural generalization of the versatile $\alpha$-$\kappa$-$\mu$ and $\alpha$-$\eta$-$\mu$ distributions. The $\alpha$-KMS fading distribution unifies a wide set of fading distributions, as it includes the $\alpha$-$\kappa$-$\mu$, $\alpha$-$\eta$-$\mu$, $\alpha$-$\mu$, Weibull, $\kappa$-$\mu$ shadowed, Rician shadowed, $\kappa$-$\mu$ and $\eta$-$\mu$ distributions as special cases, together with classical models like Rice, Nakagami-$m$, Hoyt, Rayleigh and one-sided Gaussian. 
Notably, the $\alpha$-KMS distribution reduces to a finite mixture of $\alpha$-$\mu$ distributions when the fading parameters $\mu$ and $m$ take positive integer values, so that performance analysis over $\alpha$-KMS fading channels can be tackled by leveraging previous (existing) results in the literature for the simpler $\alpha$-$\mu$ case. As application examples, important performance metrics like the outage probability and average channel capacity are analyzed.
\end{abstract}


\section{Introduction}

The $\alpha$-$\mu$ distribution originally proposed by Yacoub \cite{Yacoub2002,Yacoub2007} has become very popular in the context of wireless channel modelling, due to not only its reasonably simple analytical form but also because of its improved fit to field measurements. Its original formulation was motivated by the interest on exploring and eventually characterizing the non-linearity of the propagation medium. Interestingly, the $\alpha$-$\mu$ distribution arises as a power transformation over the received signal envelope in a Nakagami-$m$ set-up, in a similar way as the Weibull distribution is obtained from an underlying Rayleigh distribution.

The $\alpha$-$\eta$-$\mu$ and $\alpha$-$\kappa$-$\mu$ distributions \cite{Fraidenraich2006} arose as generalizations of the equally popular and versatile $\kappa$-$\mu$ and $\eta$-$\mu$ distributions \cite{Yacoub2000,Yacoub2001,Yacoub2007b}. Despite their versatility on extending the range of propagation conditions which they are able to model, their applicability for performance analysis purposes is hindered by the complicated form of their chief probability functions, and the scarce results available in the literature often have complicated form \cite{Aldalgamouni2017,Moualeu2018,Magableh2018}, compared to their $\alpha$-$\mu$ fading counterparts.

In \cite{Paris2014,Cotton2015}, the $\kappa$-$\mu$ shadowed fading distribution was introduced with the aim of generalizing the $\kappa$-$\mu$ distribution by allowing the dominant specular components to randomly fluctuate. This effect was introduced through an additional parameter $m$ similar to that in the Nakagami-$m$ distribution, and provided additional flexibility to the model in order to better accommodate to field measurements. It was later demonstrated in \cite{Moreno2016} that, somehow counterintuitively, the $\kappa$-$\mu$ shadowed fading model included as special cases \emph{both} the $\kappa$-$\mu$ and $\eta$-$\mu$ distributions. This allows to unify both fading models under a common, more general model with a similar mathematical complexity. Besides, in the special cases of considering integer values for the fading parameters $\mu$ and $m$, the probability density function (PDF) and the cumulative distribution function (CDF) of the $\kappa$-$\mu$ shadowed fading model are expressed in terms of a finite sum of powers and exponentials \cite{Lopez2017}. Hence, in these circumstances, its tractability is as simple as if the simple Nakagami-$m$ model was assumed.

As we can see from the previous works, the consideration of channel propagation effects like non-linearity of the propagation medium and the fluctuation of dominant specular components have been tackled separately to the best of our knowledge. In order to fill this gap, and therefore to jointly consider both effects, we formulate the $\alpha$-$\kappa$-$\mu$ shadowed fading model ($\alpha$-KMS) as a natural generalization of the $\kappa$-$\mu$ shadowed fading model, which inherits its main key features: (i) it includes as special cases \emph{both} the $\alpha$-$\kappa$-$\mu$ and the $\alpha$-$\eta$-$\mu$ fading models; (ii) it also includes as special cases virtually all classical models (Nakagami, Hoyt, Rice, $\kappa$-$\mu$, $\eta$-$\mu$, Weibull, $\alpha$-$\mu$, Rician Shadowed); (iii) for integer values of $\mu$ (corresponding to the physical model based on clusters in \cite{Yacoub2007b}) and $m$, the $\alpha$-KMS model can be expressed as a finite mixture of $\alpha$-$\mu$ distributions, which implies that performance analysis over $\alpha$-KMS fading channels is \emph{as tractable} as the $\alpha$-$\mu$ case. Hence, the $\alpha$-KMS fading model not only generalizes but also simplifies the analysis of the baseline fading models from which it originates. We exemplify the tractability of the $\alpha$-KMS fading channel model by analyzing two key performance metrics, i.e., the outage probability and the average channel capacity. In both cases, we provide analytical expressions for these performance metrics, together with simple but tight approximations in the high signal-to-noise ratio (SNR) regime. 

The remainder of this paper is structured as follows: the physical model for the $\alpha$-KMS fading distribution is described in Section \ref{S2}, whereas the statistical characterization of this distribution is carried out in Section \ref{S3}. The application to system performance analysis is introduced in Section \ref{S4}, covering outage probability and average channel capacity. Numerical results are presented in Section \ref{S5}, and conclusions are discussed in Section \ref{S6}.

\section{Physical Model -- $\alpha$-KMS Fading Distribution}
\label{S2}

According to the physical model of the $\kappa$-$\mu$ shadowed fading distribution originally given in \cite{Paris2014}, the received signal envelope $R_\mathcal{S}$ can be expressed as a superposition of $\mu$ multipath clusters in the following form:
\begin{equation}
	\label{eq:RS}
	R_\mathcal{S}^{2}=\sum _ { i = 1 } ^ { \mu } \left( X _ { i } + \xi p _ { i } \right) ^ { 2 } + \sum _ { i = 1 } ^ { \mu } \left( Y _ { i } + \xi q _ { i } \right) ^ { 2 }.
\end{equation}

Within each cluster, the scattered components are modeled as zero-mean independent Gaussian random variables (RVs), i.e. $X_i, Y_i\sim\mathcal{N}(0,\sigma^2)$ $\forall$ $i$, whereas the dominant components are defined by the real parameters $p_i$ and $q_i$. The RV $\xi$ is defined so that its square value is gamma distributed with unit mean and shape parameter $m$, i.e. $\xi^2\sim\Gamma(m,1/m)$, which ultimately encapsulates any sort of amplitude fluctuation suffered by the line-of-sight (LoS) component. 

The physical models of the $\alpha$-$\mu$, $\alpha$-$\kappa$-$\mu$ and $\alpha$-$\eta$-$\mu$ fading distributions are formulated by applying a power transformation over the received signal envelope, in a similar way as the Weibull distribution is derived from the Rayleigh model. Using this approach \cite{Yacoub2007}, the received signal envelope $R$ in an $\alpha$-KMS fading environment is given by  

\begin{equation}
	\label{eq:RaKMS}
R^{\alpha}=\sum _ { i = 1 } ^ { \mu } \left( X _ { i } + \xi p _ { i } \right) ^ { 2 } + \sum _ { i = 1 } ^ { \mu } \left( Y _ { i } + \xi q _ { i } \right) ^ { 2 }
\end{equation}
where $\alpha>0$ is a real number characterizing the non-linearity of the propagation medium. Note that, if $\alpha=2$ then \eqref{eq:RaKMS} reduces to \eqref{eq:RS}. The $\alpha$-KMS model is completely defined by the set of parameters $\alpha$, $\mu$, $m$ and $\kappa = \sum_i^\mu (p_i^2+q_i^2) / (2\sigma^2)$.  In the following derivations, we will pursue the statistical characterization of the RVs arising from this physical model. We will focus on the characterization of the instantaneous SNR under $\alpha$-KMS fading, defined as $\gamma=\overline\gamma R^2/ \,\mathbb{E}\left[R^2\right]$, where $\overline\gamma$ is the average SNR. From the formulations for $\gamma$, those of the received signal envelope $R$ can be easily obtained by using a simple change of variables.

\section{Statistical Characterization}
\label{S3}
\subsection{Definitions}

\begin{definition}[$\alpha$-$\mu$ distribution]
    Let $\gamma$ be a real RV characterizing the instantaneous SNR under $\alpha$-$\mu$ fading \cite{Yacoub2007b}, with mean $\overline{\gamma} = \mathbb{E}\left[\gamma\right]$ and real shape parameters $\alpha$ and $\mu$. Then, the PDF of $\gamma$ is given by
    \begin{equation}
    \label{eqAM}
        f_{\alpha\mu}(\overline{\gamma};\alpha, \mu;\gamma)= \frac{\alpha}{2\Gamma(\mu)} D^{\mu\alpha /2} \gamma^{\alpha\mu/2 -1}e^{-\gamma^\alpha D^{\alpha/2}},    \end{equation}
with $D= \frac{\Gamma\left(\mu+\frac{2}{\alpha}\right)}{\overline{\gamma} \Gamma(\mu)}$, and $\Gamma(\cdot)$ being the gamma function \cite[8.310]{Gradshteyn07}.
\end{definition}

\subsection{Chief Probability Functions}
\lemma[{The $\alpha$-KMS distribution: PDF}]{
 Let $\gamma$ be a real RV characterizing the instantaneous SNR under $\alpha$-KMS fading, with $\overline\gamma=\mathbb{E}[\gamma]$ and non-negative real shape parameters $\alpha$, $\kappa$, $\mu$ and $m$, i.e. $\gamma\sim\mathcal{A}_{kms}\left(\overline{\gamma}; \alpha,\kappa,\mu,m\right)$. Then, its PDF is given by
\begin{align}
\label{eqPDF}
f_\gamma(\gamma)=&\frac{m^m\alpha}{2c^\mu\Gamma(\mu)\left(\mu\kappa+m\right)^m \overline{\gamma}}\left(\frac{\gamma}{\overline{\gamma}}\right)^{\frac{\alpha\mu}{2}-1}\exp\left(-\frac{1}{c}\left(\frac{\gamma}{\overline{\gamma}}\right)^{\frac{\alpha}{2}}\right)\nonumber\\&\times{}_1F_{1}\left(m,\mu;\frac{\mu\kappa}{c(\mu\kappa+m)}\left(\frac{\gamma}{\overline{\gamma}}\right)^{\frac{\alpha}{2}}\right),
\end{align}
where $_1{F}_1(\cdot)$ is the confluent hypergeometric function \cite[Eq. (13.1.2)]{Abra72} and 
\begin{equation}
    c = \left(\frac{(\mu\kappa +m)^m\Gamma(\mu)}{m^m\Gamma\left(\mu+\frac{2}{\alpha}\right){}_2F_1\left(m,\mu+\frac{2}{\alpha};\mu;\frac{\mu\kappa}{\mu\kappa+m}\right)}\right)^{\alpha/2},
\end{equation}
with ${}_2F_1(\cdot)$ denoting the Gauss hypergeometric function \citep[Eq. (15.1.1)]{Abra72}.
}
\begin{IEEEproof}
Using standard techniques of transformation of random variables, and starting from the original expressions given in \cite{Paris2014,Cotton2015} for the $\kappa$-$\mu$ shadowed fading distribution, the PDF of the signal amplitude $R$ can be deduced as
\begin{align}
f_R(r)=&\frac{m^m\alpha r^{\alpha\mu-1}}{\Omega^{\alpha\mu/2}c^\mu\Gamma(\mu)\left(\mu\kappa+m\right)^m}\exp\left(-\frac{r^\alpha}{\Omega^{\alpha/2}c}\right)\nonumber\\ &\times{}_1F_{1}\left(m,\mu;\frac{\mu\kappa r^\alpha}{\Omega^{\alpha/2}c(\mu\kappa+m)}\right),
\end{align}
where $\Omega=\E\left[R^2\right]$. Finally, since $\gamma=\overline\gamma\frac{R^2}{\Omega}$ we can obtain the desired PDF as $f_{\gamma}(\gamma)=\frac{1}{2}\sqrt{\frac{\Omega}{\overline\gamma\gamma}}f_{R}\left(\sqrt{\frac{\Omega\gamma}{\overline\gamma}}\right)$.
\end{IEEEproof}
\lemma[{The $\alpha$-KMS distribution: CDF}]{
 Let $\gamma$ be a real RV characterizing the instantaneous SNR under $\alpha$-KMS fading, with $\overline\gamma=\mathbb{E}[\gamma]$ and non-negative real shape parameters $\alpha$, $\kappa$, $\mu$ and $m$, i.e. $\gamma\sim\mathcal{A}_{kms}\left(\overline{\gamma}; \alpha,\kappa,\mu,m\right)$. Then, its CDF is given by
\begin{align}
\label{eqCDF}
& F _ { \gamma } ( \gamma ) = \frac {m ^ { m } } {\mu c^\mu \Gamma ( \mu ) ( \mu \kappa + m ) ^ { m } }  \left( \frac { \gamma } { \overline { \gamma } } \right) ^ { \alpha\mu/2 } \\
 &\times  { \Phi _ { 2 } \left( \mu - m , m , \mu + 1 ; - \tfrac{1}{c} \left(\tfrac { \gamma } { \overline { \gamma } }\right)^{\frac{\alpha}{2}} , - \left(\tfrac { \gamma  } { \overline { \gamma } }\right)^{\frac{\alpha}{2}} \tfrac { m } {c(\mu \kappa + m) } \right), \nonumber} \label{eq:CDF}
\end{align}
where $\Phi_2$ is the confluent bivariate hypergeometric function defined in \cite[Eq. (7.2)]{Srivastava1985}.
}
\begin{IEEEproof}
Following the same steps as in the previous proof, using now as a starting point the CDF of $R$ in \cite{Paris2014,Cotton2015}, we obtain the desired result.
\end{IEEEproof}




\lemma[{The $\alpha$-KMS distribution: Moments}]{
 Let $\gamma$ be a real RV characterizing the instantaneous SNR under $\alpha$-KMS fading, with $\overline\gamma=\mathbb{E}[\gamma]$ and non-negative real shape parameters $\alpha$, $\kappa$, $\mu$ and $m$, i.e. $\gamma\sim\mathcal{A}_{kms}\left(\overline{\gamma}; \alpha,\kappa,\mu,m\right)$. Then, the $n^{th}$ moment of the SNR is given by
\begin{equation}
	\mathbb{E}\left[\gamma^n\right] = \overline{\gamma}^n\frac{m^m\Gamma\left(\mu + n\frac{2}{\alpha}\right)c^{\frac{2n}{\alpha}}}{\Gamma(\mu)(\mu\kappa + m)^m}{}_2F_1\left(m,\mu+n\frac{2}{\alpha};\mu;\tfrac{\mu\kappa}{\mu\kappa + m}\right). \label{eq:Moments}
\end{equation}
}
\begin{IEEEproof}
Using \eqref{eqPDF} and \cite[Eq. (7.621 4)]{Gradshteyn07} yields the desired result.
\end{IEEEproof}

Expressions \eqref{eqPDF}, \eqref{eqCDF} and \eqref{eq:Moments} are new in the literature to the best of our knowledge, and their validity has been extensively corroborated using Monte Carlo simulations. Direct inspection of the chief statistics of the newly proposed fading model reveals that they pose no additional challenges from a computational perspective that those of the $\kappa$-$\mu$ shadowed fading case \cite{Paris2014,Cotton2015}. We note that the confluent bivariate hypergeometric function $\Phi_2(\cdot)$ also appears in state-of-the-art fading models \cite{Morales2010,Paris2010,Romero2017}, and can be efficiently evaluated \cite{Martos2016}. The connections between the $\alpha$-KMS distribution and other popular fading models are summarized in Table \ref{Table1}. Notably, the $\alpha$-$\kappa$-$\mu$ and the $\alpha$-$\eta$-$\mu$ distributions proposed in \cite{Fraidenraich2006} from two different underlying physical models are now expressed as special cases of the $\alpha$-KMS distribution.
 
 \begin{table}[t]
\renewcommand{\arraystretch}{1.7}
\caption{Connections between the $\alpha$-KMS fading model and other models in the literature.}
\label{Table1}
\centering
\begin{tabular}
{c|c}
\hline
\hline
Channels  & $\alpha$-KMS Fading Parameters\\
\hline
\hline
\multirow{1}{*}{One-sided normal} &  $\underline{\alpha}=2$,\hspace{3mm}$\underline{\kappa}=0$,\hspace{3mm}$\underline{\mu}=1/2$,\hspace{3mm}$\forall\;\underline{m}$\\
\hline
\multirow{1}{*}{Rayleigh} &  $\underline{\alpha}=2$,\hspace{3mm}$\underline{\kappa}=0$,\hspace{3mm}$\underline{\mu}=1$,\hspace{3mm} $\forall\;\underline{m}$\\
\hline
\multirow{1}{*}{Nakagami-$m$} & $\underline{\alpha}=2$,\hspace{3mm}$\underline{\kappa}=0$,\hspace{3mm}$\underline{\mu}=m$,\hspace{3mm}$\forall\;\underline{m}$ \\
\hline
\multirow{1}{*}{Hoyt}  & $\underline{\alpha}=2$,\hspace{3mm}$\underline{\kappa}=(1-q^2)/(2q^2)$,\hspace{3mm}$\underline{\mu}=1$,\hspace{3mm} $\underline{m}=1/2$\hspace{3mm}\\
\hline
\multirow{1}{*}{$\eta$-$\mu$}& $\underline{\alpha}=2$,\hspace{3mm}$\underline{\kappa}=(1-\eta)/(2\eta)$,\hspace{3mm}$\underline{\mu}=2\mu$,\hspace{3mm} $\underline{m}=\mu$\hspace{3mm}\\
\hline
Rice &  $\underline{\alpha}=2$,\hspace{3mm}$\underline{\kappa}=K$,\hspace{3mm}$\underline{\mu}=1$\hspace{3mm}, $\underline{m}\to\infty$\hspace{3mm}\\
\hline
$\kappa$-$\mu$ &  $\underline{\alpha}=2$,\hspace{3mm}$\underline{\kappa}=\kappa$,\hspace{3mm}$\underline{\mu}=\mu$\hspace{3mm}, $\underline{m}\to\infty$\hspace{3mm}\\
\hline
Rician shadowed &  $\underline{\alpha}=2$,\hspace{3mm}$\underline{\kappa}=K$,\hspace{3mm}$\underline{\mu}=1$\hspace{3mm}, $\underline{m}=m$\hspace{3mm}\\
\hline
$\kappa$-$\mu$ shadowed &  $\underline{\alpha}=2$,\hspace{3mm}$\underline{\kappa}=\kappa$,\hspace{3mm}$\underline{\mu}=\mu$\hspace{3mm}, $\underline{m}=m$\hspace{3mm}\\
\hline
Weibull&  $\underline{\alpha}=\alpha$,\hspace{3mm}$\underline{\kappa}=0$,\hspace{3mm}$\underline{\mu}=1$\hspace{3mm}, $\forall\;\underline{m}$\hspace{3mm}\\
\hline
$\alpha$-$\mu$&  $\underline{\alpha}=\alpha$,\hspace{3mm}$\underline{\kappa}=0$,\hspace{3mm}$\underline{\mu}=\mu$\hspace{3mm}, $\forall\;\underline{m}$\hspace{3mm}\\
\hline
$\alpha$-$\kappa$-$\mu$& $\underline{\alpha}=\alpha$,\hspace{3mm}$\underline{\kappa}=\kappa$,\hspace{3mm}$\underline{\mu}=\mu$\hspace{3mm}, $\underline{m}\to\infty$\hspace{3mm}\\
\hline
$\alpha$-$\eta$-$\mu$& $\underline{\alpha}=\alpha$,\hspace{3mm}$\underline{\kappa}=(1-\eta)/(2\eta)$,\hspace{3mm}$\underline{\mu}=2\mu$,\hspace{3mm} $\underline{m}=\mu$\hspace{3mm}\\
\hline
\end{tabular}
\end{table} 

 \subsection{Integer values of $\mu$ and $m$}
 
 As with the original $\kappa$-$\mu$ shadowed fading distribution \cite{Lopez2017}, the expressions for the PDF and CDF of the $\alpha$-KMS model are considerably simplified when considering both $m$ and $\mu$ parameters to be positive integer numbers. Thus, using again standard techniques of transformation of random variables from the original expressions in \cite{Lopez2017}, simplified expressions for the PDF and CDF are given in the following lemma.
 
 \lemma[{The $\alpha$-KMS distribution with integer $\mu$ and $m$}]{
 Let $\gamma$ be a real RV characterizing the instantaneous SNR under $\alpha$-KMS fading, with $\overline\gamma=\mathbb{E}[\gamma]$ and non-negative real shape parameters $\alpha$, $\kappa$, and integer shape parameters $\mu$ and $m$. Then, the PDF and the CDF of $\gamma$ can be expressed in closed-form as
 \begin{equation}
     f_{\gamma}(\gamma) = \sum_{i=0}^M C_i \frac{\alpha}{2\left(\overline{\gamma}^{\alpha/2}c B_i\right)^{m_i}} \frac{\gamma^{\alpha m_i/2-1}}{(m_i-1)!} \exp\left(-\frac{\gamma^{\alpha/2}}{\overline{\gamma}^{\alpha/2} cB_i}\right),  \label{eq:PDFinteger}
 \end{equation}
 \begin{equation}
     F_\gamma(\gamma) = 1-\sum_{i=0}^M C_i \exp\left(-\frac{\gamma^{\alpha/2}}{\overline{\gamma}^{\alpha/2}cB_i}\right)\sum_{j=0}^{m_i-1} \frac{1}{j!}\left(\frac{\gamma^{\alpha/2}}{\overline{\gamma}^{\alpha/2}cB_i}\right)^j,
 \end{equation}
  where the parameters $M$, $C_i$, $m_i$ and $B_i$ are expressed in Table \ref{table01} in terms of the parameters $\kappa$, $\mu$ and $m$. 
}
\begin{IEEEproof}
Using the same steps as in the derivation of \eqref{eqPDF} and \eqref{eqCDF}, now using as starting points the simplified PDF and CDF for the case of $m$ and $\mu$ being integers given in \cite{Lopez2017}, the proof is completed.
\end{IEEEproof}

 
  \begin{table*}[t]
  \renewcommand{\arraystretch}{3}
\centering
{
\caption{Parameter values for the $\alpha$-KMS distribution with integer $\mu$ and $m$,}
\label{table01}
\begin{tabular}{|c|c|}
\hline\hline
Case $\mu>m$ & Case $\mu \leq m$ \\ \hline\hline 
$M=\mu$ & $M=m-\mu$  \\ \hline
 $C_i=\begin{cases} 
      0 & i=0 \\
       \left( { - 1} \right)^m \binom{m+i-2}{i-1}\times \left[ {\frac{m}
{{\mu \kappa  + m}}} \right]^{ m} \left[ {\frac{{\mu \kappa }}
{{\mu \kappa  + m}}} \right]^{ - m - i + 1}  & 0<i\leq \mu-m \\
      \left( { - 1} \right)^{i-\mu+m - 1} \binom{i-2}{i-\mu+m-1} \times \left[ {\frac{m}
{{\mu \kappa  + m}}} \right]^{i-\mu+m - 1} \left[ {\frac{{\mu \kappa }}
{{\mu \kappa  + m}}} \right]^{-i + 1}  & \mu-m < i \leq \mu 
   \end{cases}$
 & $C_i=\binom{m-\mu}{i}\left[ {\frac{m}
{{\mu \kappa  + m}}} \right]^i \left[ {\frac{{\mu \kappa }}
{{\mu \kappa  + m}}} \right]^{m - \mu  - i}$  \\ \hline
\ $m_i=\begin{cases} \mu-m-i+1, & 0\leq i\leq \mu-m \\
   \mu-i+1 & \mu-m < i \leq \mu 
   \end{cases}$ & $m_i=m-i$  \\ \hline
    $B_i=\begin{cases}  1, & 0\leq i\leq \mu-m \\
   \frac{{\mu \kappa  + m}}
{m} & \mu-m < i \leq \mu 
   \end{cases}$ & $B_i=\frac{{\mu \kappa  + m}}
{m}$  \\ \hline
\end{tabular}
}
\end{table*}
%
 
We observe that the PDF and the CDF in this case are expressed in terms of a finite sum of powers and exponential functions. This greatly simplifies the numerical evaluation of the statistics of the $\alpha$-KMS distribution, as well as facilitates subsequent mathematical derivations for performance analysis purposes. Taking a deeper look at the expressions in Lemma 4, and using the definition for the $\alpha$-$\mu$ distribution in \eqref{eqAM}, we see that the $\alpha$-KMS distribution can be seen as a finite mixture of $\alpha$-$\mu$ distributions, as
 \begin{equation}
 \label{mixture}
 	f_{\gamma}(\gamma) = \sum_{i=0}^M C_i f_{\alpha\mu}(\omega_i; \alpha, m_i, \gamma)
 \end{equation}
 where 
 \begin{equation}
 	\omega_i = \frac{\overline{\gamma}(cB_i)^{2/\alpha}\Gamma(m_i+2/\alpha)}{\Gamma(m_i)}.
 \end{equation}

Because of this finite mixture representation for the $\alpha$-KMS distribution, its PDF is given as a weighted sum of $\alpha$-$\mu$ PDFs. This implies that any system performance metric for the $\alpha$-KMS fading channel (with integer $\mu$ and $m$) can be directly obtained from previous (existing) results for the conventional $\alpha$-$\mu$ case, without the need to resort to further analytical manipulations.

 \section{Applications}
\label{S4}
\subsection{Outage probability}

The outage probability is defined as the probability that the instantaneous SNR falls below a predefined threshold value $\gamma_{\rm th}$, i.e., $P_{\rm out}\triangleq \Pr\{\gamma<\gamma_{\rm th}\}$. Hence, the OP under $\alpha$-KMS fading can be directly obtained by evaluating the CDF of the SNR in \eqref{eqCDF} as $P_{\rm out}=F_{\gamma}(\gamma_{\rm th})$.

In the high-SNR regime, i.e., $\overline{\gamma}\to\infty$, the $\Phi_2(\cdot)$ function in \eqref{eqCDF} tends to one, so the asymptotic OP can be approximated as
\begin{equation}
\label{eqCDF2}
	\left.P_{\rm out}\right|_{\overline{\gamma}\Uparrow} \approx \frac {m ^ { m } } {\mu c^\mu \Gamma ( \mu ) ( \mu \kappa + m ) ^ { m } }  \left( \frac { \gamma_{th} } { \overline { \gamma } } \right) ^ { \alpha\mu/2 }.
\end{equation}

Direct inspection of \eqref{eqCDF2} reveals that the diversity order of the $\alpha$-KMS fading model (i.e., the asymptotic decay of the OP) is $d=\alpha\mu/2$.
\subsection{Average channel capacity}

The average channel capacity per unit bandwidth is defined as
\begin{equation}
\label{capacidad}
\overline C(\overline\gamma)\triangleq \int_{0}^{\infty} \log_2\left(1+\gamma\right)f_{\gamma}(\gamma),
\end{equation}
i.e., the instantaneous Shannon capacity averaged over all possible fading states. An integral expression for this capacity can be obtained by plugging \eqref{eqPDF} into \eqref{capacidad}. In the specific case of $\mu$ and $m$ being integers, the capacity can be directly obtained from that of the $\alpha$-$\mu$ case using \eqref{mixture} and the results in \cite[Eq. (2)]{Costa2007}.

 In the high-SNR regime, the average capacity can be approximated by \cite[Eq. (8)]{Yilmaz2012}
\begin{equation}
	\left.\overline{C}(\overline{\gamma})\right|_{\overline{\gamma}\Uparrow} = {\rm log}_2(\overline{\gamma}) - L
\end{equation}
where
\begin{align}
	L &= - {\rm log}_2 (e)  \left.\frac{d}{dn}\frac{\E\{\gamma^n\}}{\overline\gamma^n}\right|_{n=0} \nonumber\\
	&= - {\rm log}_2 (e)  \left.\frac{d}{dn}G(n)\right|_{n=0} = - {\rm log}_2 (e) G'(0).
\end{align}

From \cite[Eq. (3)]{Yilmaz2012} and \eqref{eq:Moments} and applying the derivative product rule, we have
\begin{equation}
	 G'(0)=\frac{2}{\alpha}\left(\psi(\mu) + {\rm ln}(c)\right) + \frac{m^m}{(\mu\kappa+m)^m}\left.\frac{\partial P(n)}{\partial n} \right|_{n=0},
\end{equation}
with $P(n)={}_2F_1\left(m,\mu+n\frac{2}{\alpha};\mu; \frac{\mu\kappa}{\mu\kappa + m}\right)$ and where $\psi(\cdot)$ is the digamma function \cite[Eq. (6.3.1)]{Abra72}. Finally, using the result for the derivative of the Gauss hypergeometric function given in \cite[app. B]{Moreno2016} the asymptotic capacity admits the following expression:
\begin{align}
\label{capacidad2}
&	\left.\overline{C}(\overline{\gamma})\right|_{\overline{\gamma}\Uparrow} = {\rm log}_2\overline{\gamma} + \frac{2}{\alpha}{\rm log}_2e \left[\psi(\mu) + {\rm ln}c -{\rm ln}\left(\frac{m}{\mu\kappa+m}\right) \right.\notag \\
	&\left. - \frac{\kappa(\mu-m)}{\mu\kappa+m}{}_3F_2\left(\mu-m+1,1,1;\mu+1,2;\frac{\mu\kappa}{\mu\kappa+m}\right)\right],
\end{align}
where ${}_3F_2(\cdot)$ is a generalized hypergeometric function \cite[p. 19]{Srivastava1985}.
 
\section{Numerical Results}
\label{S5}
\subsection{Effect of fading parameters}
 
 We now exemplify how the fading parameters of the $\alpha$-KMS fading model affect the shape of the distribution. For the sake of compactness and because of space limitations, we focus on the PDF of the SNR.

\begin{figure}[t]
    \centering
     \includegraphics[width=1.00\columnwidth]{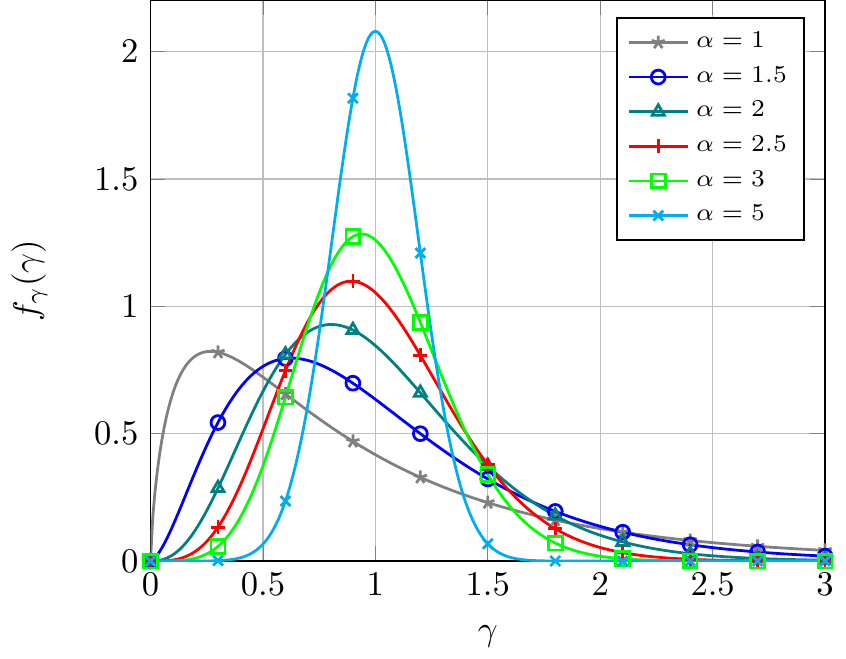}
         \caption{Evaluation of the $\alpha$-KMS fading PDF for different values of $\alpha$, with $\overline\gamma=1$ (i.e., 0 dB). Parameter values are $\kappa=3$, $\mu=3.2$ and $m=7.3$.}
      \label{fig:0}
\end{figure}

The evolution of the PDF in \eqref{eqPDF} as the shape parameter $\alpha$ varies is represented in Fig. \ref{fig:0}, where the rest of parameters are set to fixed values, i.e., $\overline\gamma=1$ (i.e. $0$ dB), $\kappa=3$, $\mu=3.2$ and $m=7.3$. We see that as $\alpha$ is increased, the probability of having very low SNR values is reduced and the SNR tends to be more concentrated around its mean value $\overline\gamma$ .

\begin{figure}[t]
    \centering
     \includegraphics[width=1.00\columnwidth]{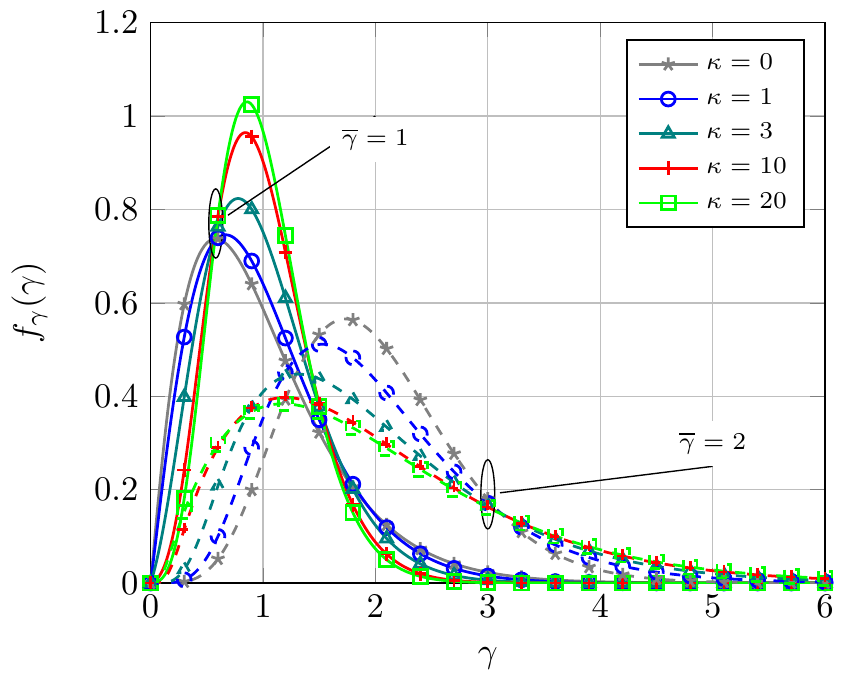}
         \caption{Evaluation of the $\alpha$-KMS fading PDF for different values of $\kappa$. Solid lines correspond to the $m>\mu$ case, with $\overline\gamma=1$ (i.e., 0 dB) and parameter values $\alpha=2.3$, $\mu=1.8$ and $m=5.5$. Dashed lines correspond to the $m<\mu$ case, with $\overline\gamma=2$ (i.e., 3 dB) and parameter values $\alpha=2.3$, $\mu=5.5$ and $m=1.8$. }
      \label{fig:1}
\end{figure}

The effect of the parameter $\kappa$ is now investigated in Fig. \ref{fig:1}. We analyze the cases on which $m>\mu$ and $m<\mu$ separately. Besides, in order for these PDFs not to be overlapped in the figure, we set different values of $\overline\gamma$ for each situation whereas $\alpha=2.3$ in all cases. Inspecting the solid line curves in Fig. \ref{fig:1}, we see that when $m>\mu$ increasing $\kappa$ reduces the fading severity in the sense that low SNR values are less likely. However, the converse behavior is observed when $m<\mu$. In this case, because the fluctuations of the dominant components are larger than those of the diffuse components, the fading severity is increased as $\kappa$ grows. We note that similarly to the $\kappa$-$\mu$ shadowed case \cite{Moreno2016}, setting $\mu=m$ implies that the parameter $\kappa$ has no effect over the shape of the distribution. Hence, for the sake of compactness this latter comparison is not included in the figure.

\begin{figure}[t]
    \centering
     \includegraphics[width=0.99\columnwidth]{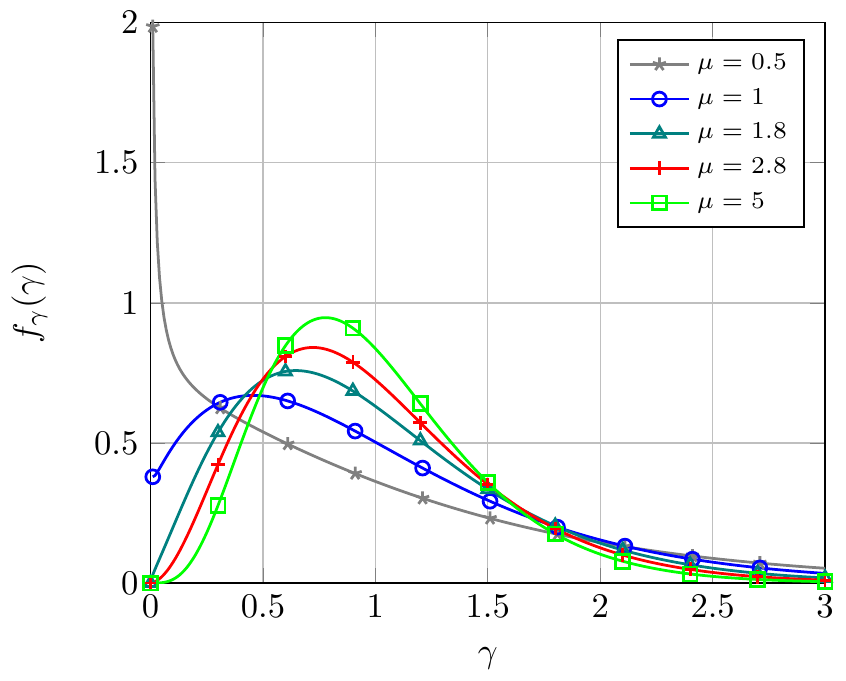}
         \caption{Evaluation of the $\alpha$-KMS fading PDF for different values of $\mu$, with $\overline\gamma=1$ (i.e., 0 dB). Parameter values are $\alpha=1.8$, $\kappa=4$ and $m=6$.}
      \label{fig:2}
\end{figure}

In Fig. \ref{fig:2} we study the effect of changing the parameter $\mu$ over the shape of the PDF, for $\overline\gamma=1$, $\kappa=4$, $\alpha=1.8$ and $m=6$. We see that increasing $\mu$, i.e., increasing the number of clusters in \eqref{eq:RaKMS}, is beneficial in terms of fading severity and hence low SNR values have lower probability as $\mu$ grows.

\begin{figure}[t]
    \centering
     \includegraphics[width=1.00\columnwidth]{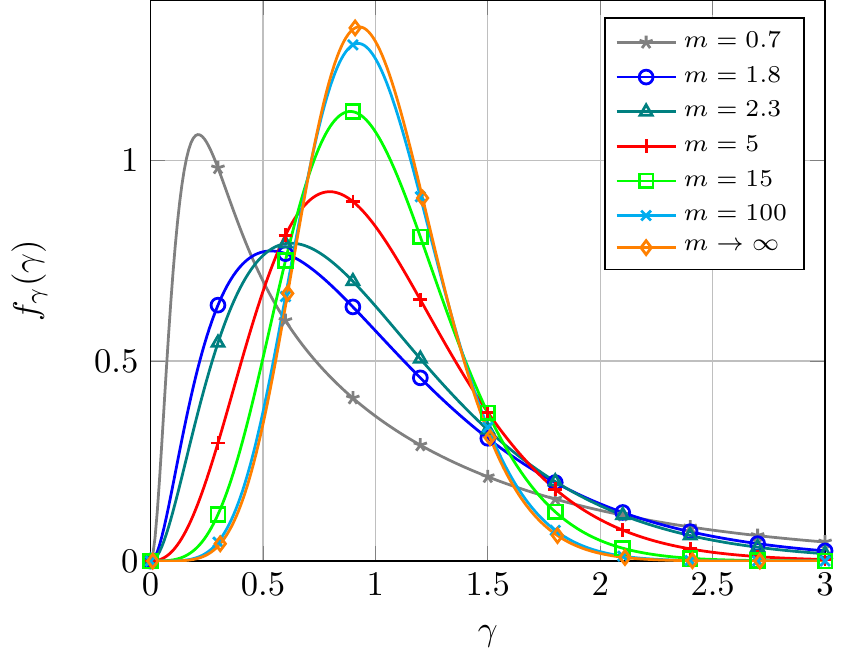}
         \caption{Evaluation of the $\alpha$-KMS fading PDF for different values of $m$, with $\overline\gamma=1$ (i.e., 0 dB). Parameter values are $\alpha=2.1$, $\kappa=5$ and $\mu=3$.}
      \label{fig:3}
\end{figure}

Finally, the effect of the parameter $m$ is depicted in Fig. \ref{fig:3}, with the rest of parameters being set to $\alpha=2.1$, $\kappa=5$ and $\mu=3$. We observe that lower values of $m$ are translated into a larger probability of having low SNR values, i.e. a larger fading severity. As $m$ grows, the effect of LOS fluctuations vanishes and the SNR values tend to be more concentrated around $\overline\gamma=1$. For sufficiently large values of $m$, the $\alpha$-KMS distribution tends to overlap with the $\alpha$-$\kappa$-$\mu$ one, for which the limit case of $m\rightarrow\infty$ is also represented in the figure. We note that the convergence in distribution between the $\alpha$-KMS distribution and the $\alpha$-$\kappa$-$\mu$ distribution as $m\rightarrow\infty$ can be formally proven using Levy's continuity theorem as in \cite{Lopez2017}.

\subsection{Performance analysis}

\begin{figure}[t]
    \centering
     \includegraphics[width=1.00\columnwidth]{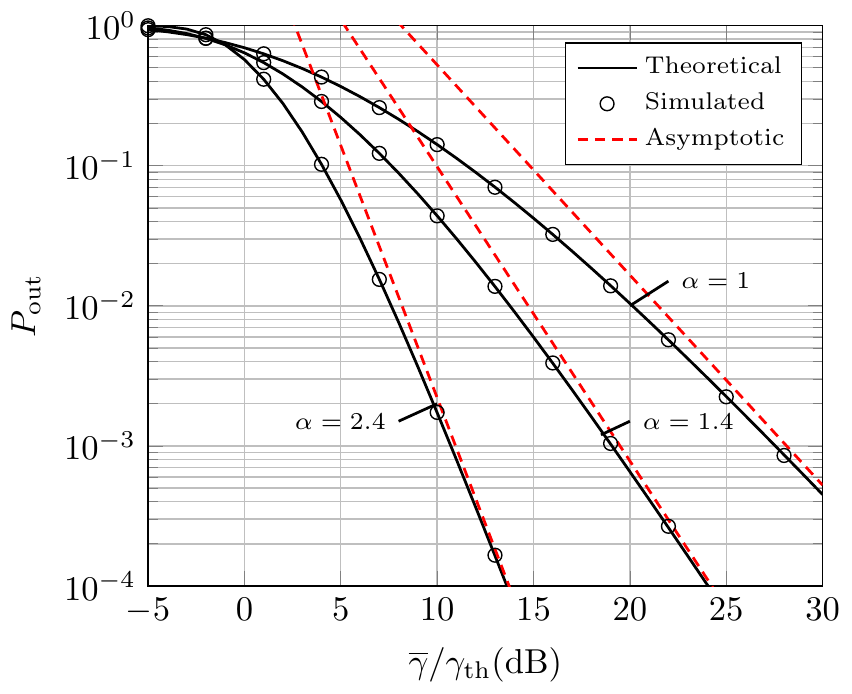}
         \caption{Outage probability vs. normalized $\overline{\gamma}$ for $\alpha$-KMS fading with $\kappa = 3.1$, $\mu = 3$, $m = 2$ and different values of $\alpha$.}
      \label{fig:4}
      \vspace{-2mm}
\end{figure}

In Fig. \ref{fig:4}, the OP under $\alpha$-KMS fading is evaluated as a function of the average SNR (normalized to the threshold value $\gamma_{th}$), for different values of the fading parameter $\alpha$. Exact and approximate expressions for the OP are computed by evaluating \eqref{eqCDF} and \eqref{eqCDF2}, respectively. Monte Carlo simulations (with markers) are superimposed to double-check the validity of the theoretical expressions. We see that the OP is improved as $\alpha$ is increased, and that the asymptotic OP values are tight for sufficiently large $\overline\gamma/\gamma_{th}$. We also see that the OP curves exhibit different diversity orders as $\alpha$ changes.

\begin{figure}[t]
    \centering
     \includegraphics[width=1.00\columnwidth]{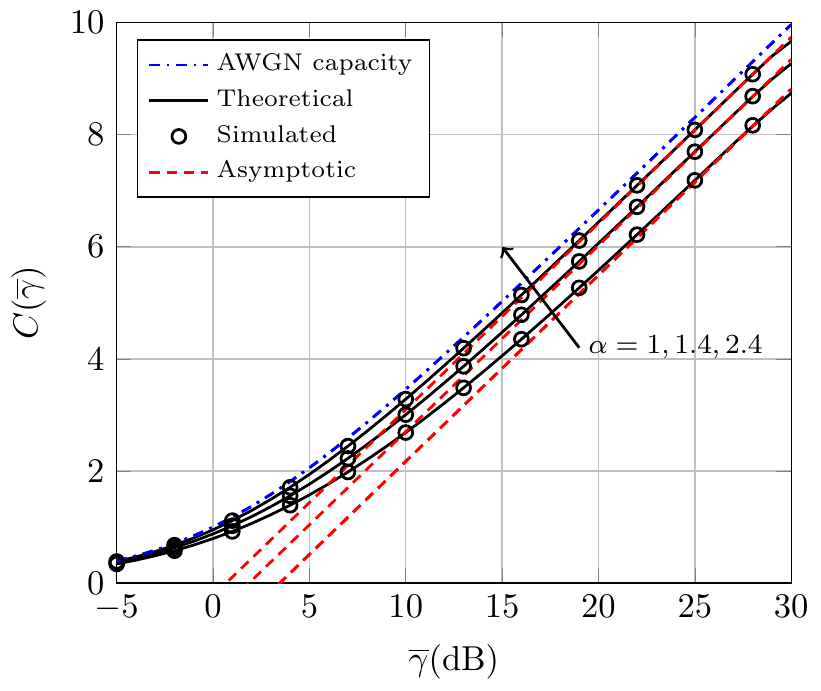}
         \caption{Average capacity vs. $\overline{\gamma}$ for $\alpha$-KMS fading with $\kappa = 3.1$, $\mu = 3$, $m = 2$ and different values of $\alpha$.}
      \label{fig:5}
      \vspace{-2mm}
\end{figure}

In Fig. \ref{fig:5}, the average capacity is analyzed for different values of the fading parameter $\alpha$, by using \eqref{capacidad} and \eqref{capacidad2} together with Monte Carlo simulations. The capacity of the AWGN case is also included as a reference value, as it always upperbounds the capacity of a fading channel due to Jensen's inequality. We see that as $\alpha$ is reduced, the capacity is also reduced. Conversely, as $\alpha$ grows, the performance gap with respect to the AWGN case is reduced.

\section{Conclusion}
\label{S6}
We introduced the $\alpha$-KMS fading model for the first time in the literature, which is strongly motivated by two main reasons: unification of two families of apparently unrelated fading models ($\alpha$-$\kappa$-$\mu$ and $\alpha$-$\eta$-$\mu$), and analytical tractability, as it reduces to a finite mixture of the much simpler $\alpha$-$\mu$ distribution for the case of $\mu$ and $m$ being integers. The performance analysis of wireless communication systems operating over this channel was exemplified for two key metrics like the OP and the average capacity. 

\section*{Acknowledgements}
This work has been supported by the Spanish Government (Ministerio de Economia y Competitividad) under grant TEC2017-TEC2017-87913-R, and by Universidad de Malaga - Campus de Excelencia Internacional Andalucia Tech.

\bibliographystyle{IEEEtran}
\bibliography{Bibliografia}

\end{document}